\documentstyle[multicol,aps,epsfig]{revtex}

\begin{document}

\def\dd{{\rm d}}
\def\ba{\begin{eqnarray}}
\def\ea{\end{eqnarray}}
\def\bqo{\begin{quote}}
\def\eqo{\end{quote}}
\def\bc{\begin{center}}
\def\ec{\end{center}}
\def\om{\omega}
\def\xs{\xi^s}
\def\ep{\epsilon}
\def\vaep{\varepsilon}
\def\re{{\rm Re}}
\def\im{{\rm Im}}
\def\cii{\chi_{\rm{imp}}}
\def\ciif{\chi_{\rm{imp1}}}
\def\ciis{\chi_{\rm{imp2}}}
\def\tp{\tilde{p}}
\def\tw{\tilde{\omega}}
\def\Res{{\rm Residues}}
\def\sgn{{\rm sgn}}
\def\ip{ip_{n}}

\title{ Optical conductivity in the normal state 
fullerene superconductors}

\author{Jung-Woo Yoo and Han-Yong Choi}

\address{ Department of physics, 
Sung Kyun Kwan University, Suwon 440-746, Korea}

\date{\today}

\maketitle

\begin{abstract}
We calculate the optical conductivity, $\sigma(\omega)$, 
in the normal state fullerene superconductors
by self-consistently including the impurity scatterings, the electron-phonon and 
electron-electron Coulomb interactions.
The finite bandwidth of the fullerenes is explicitely considered, and the vertex corection
is included $a$ $la$ Nambu in calculating the renormalized Green's function.
$\sigma(\omega)$ is obtained by calculating the current-current correlation 
function with the renormalized Green's function in the Matsubara frequency 
and then performing analytic continuation to the real frequency 
at finite temperature. 
The Drude weight in $\sigma(\omega)$ is strongly suppressed 
due to the interactions 
and transfered to the mid-infrared region around and above $0.06$ eV 
which is somewhat less pronounced and much broader compared with the
expermental observation by DeGiorgi $et$ $al$. 

\end{abstract}


\pacs{PACS numbers: 74.70.Wz, 74.20.Fg, 74.25.Nf}

\begin{multicols}{2}

\section{Introduction}

The optical spectra of fullerene superconductors in the normal state
were found to exhibit some unusual features \cite{degiorgi,iw}.
The optical conductivity, $\sigma(\omega)$, deviates considerably from the simple Drude behavior
expected for conventional metals: the spectral weight of 
the Drude peak is reduced by about an order of magnitude and transfered
to a mid-infrared (MIR) region around $0.06$ eV.
This suggests that the strong interaction effects due to the
Coulomb and electron-phonon interactions should be important in 
the optical spectra of the fullerenes. 
Understanding this unusual behavior in the optical conductivty,
therefore, could reveal important information about 
the fullerenes and 
contribute to understanding other physical properties of the material.

The optical conductivity $\sigma(\omega)$ represents the rate at which 
electrons absorb the incident photons at energy $\omega$, 
and is a useful probe in determining electronic characteristics of the material under study.
For an ideal free electron gas, where the interactions between the electrons, and
between the electrons and phonons are neglected, and the impurity
scattering rate $1/\tau \to 0$, 
$\sigma(\omega)$ collapses to a delta-function, 
$\sigma(\omega) = D_{tot}\delta(\omega)$,
where the coefficient $D_{tot}$ represents the total spectral weight. 
In this case, the optical conductivity sum rule,
$\int_0^{\infty} d\omega \sigma(\omega) = D_{tot}=\pi e^2 n/m$,
where $n$ is the density and $m$ is the mass of the electrons,
is exhausted entirely by the delta function Drude contribution alone.
When the material becomes dirtier, 
the Drude peak of the optical conductivity acquires the Lorentzian shape 
with the width of $1/\tau$.
The conductivity sum rule is still exhausted by the Drude part alone 
when only the impurity scatterings are present in the system.
The total weight $D_{tot}$, however, can change as the inpurity scatterings or 
other interactions are introduced when we consider a finite bandwidth, 
because the projection to a restricted basis set disregards all excitations 
to higher energy than the bandwidth. 
When other interactions are present, the free-carrier 
Drude weight is reduced by the quasiparticle renormalization
factor $Z$ such that $D = Z^{-1}D_{tot}$, and the missing spectral weight from the Drude part is
transfered to a higher energy region of $\sigma(\omega)$ reflecting the 
excitation of incoherent scatterings.

The experimentally measured $\sigma(\omega)$ in the normal state $\rm{A_3 C_{60}}$ shows 
a remarkable reduction of Drude weight and, 
concomitantly, a pronounced MIR absorption below the inter-band absorption peak:
DeGiorgi $et ~al.$ found a pronounced MIR peak around 0.06 eV and analysized that the Drude weight 
is reduced to about $0.1 - 0.2$ of the total intra-band spectral 
weight \cite{degiorgi},
while Iwasa $et ~al.$ observed the MIR absorption peak
around 0.4 eV and determined that the Drude weight is reduced
to about 0.6 of the total intra-band spectral weight \cite{iw}.
Although their results show somewhat different Drude weight and MIR absorption energy each other,
the pronounced suppression of the Drude weight and the accompanying MIR absorption 
imply strong electron-phonon and/or electron-electron interactions in this material.

In order to understand this unusual feature in $\sigma(\omega)$ of doped fullerenes,
Gunnarsson $et~ al.$ studied the effects of the electron-phonon interaction on $\sigma(\omega)$
assuming that the Migdal theorem is valid \cite{gu}.
They showed that the electron-phonon interaction leads to a narrowing of the Drude peak 
by the factor $Z=1+\lambda$, where $\lambda$ is the dimensionless electron-phonon coupling constant, 
and a transfer of the depleted Drude weight to a MIR region
at somewhat larger energies than the phonon energy. Their results, however, are
far from sufficient to describe experimental observations.
Therefore, they hinted that the Coulomb interaction
between conduction electrons, which is neglected in their study, could
lead to futher reduction of the Drude weight and more pronounced MIR absorption.
On the other hand, one of the present authors recently found, 
by studying the NMR coherence peak supression 
in the fullerene superconductors, that the Coulomb interaction between conduction electrons,
characterized by $UN_F \approx 0.3 - 0.4$, where $U$ is the effective Coulomb interaction
and $N_F$ is the density of states (DOS) at the Fermi level, should be included in 
addition to the electron-phonon interaction to understand the various experimental observations
in fullerenes in a coherent way \cite{pro1}. 
We, therefore, included the electron-electron as well as electron-phonon interactions 
at the presence of the impurity scatterings in the present paper, to better 
understand the experimentally obsered unusual features in the optical spectra
of the fullerene superconductors in the normal state.

For fullerene superconductors, the Fermi 
energy $\varepsilon_F = B/2 \approx 0.2-0.3$ eV and the average phonon frequency 
$\omega_{ph} \approx 0.05-0.15$ eV, where $B$ is the bandwidth. 
Therefore, $\omega_{ph}/\varepsilon_F \sim 1$
for fullerenes unlike conventional metals, where $\omega_{ph}/\varepsilon_F \ll 1$.
When $\omega_{ph}/\varepsilon_F \sim 1$, the phonon vertex correction 
becomes important because the Migdal theorem does not hold \cite{mi,sc,al},
and the frequency dependence of the effective Coulomb interaction, $V_{eff}(\omega)$, 
should be considered because the frequency scale at which $V_{eff} (\omega)$ varies
is comparable with that of electron-phonon interaction \cite{gu2}.
In this present work, concerned with the effects of the Coulomb and electron-phonon interactions
on the optical spectra in the narrow band fullerene superconductors,
the vertex correction is incorporated in calculating the electron self-energy
\cite{pro1,na,pro2}. The Coulomb interaction, modelled in terms of the 
onsite Hubbard repulsion, is included on an equal footing with the electron-phonon interaction,
and considered fully self-consistently in calculating the effective electron-electron interaction
\cite{pro1,pro2}.
The effective electron-electron interaction
becomes frequency dependent through the screening. The impurity effects are included 
with the $t$-matrix approximation. 

Through the relation 
\ba
\Sigma(ip) = G^{-1}(ip) - G_{0}^{-1}(ip),
\label{s1}
\ea
one obtain the electron self-energy $\Sigma(ip)$ in the Mastubara frequency, 
which gives $\Sigma(\omega)$ in the real frequency after the analytic continuation.
$G_0$ and $G$ are, respectively, the bare and renormalized electron Green's functions.
$\Sigma(\omega)$ or $Z(\omega)$, where the renormalization function $Z(\omega)$
is given by $\Sigma(\omega)=\omega-\omega Z(\omega)$, defines the single-particle
Green's function of an interacting system as 
\ba
G^{-1}=\omega-\xi_k-\Sigma(\omega)=\omega Z(\omega)-\xi_k,
\label{s2}
\ea
where $\xi_k$ is the electron energy measured from the chemical potential,
$\xi_k = \varepsilon_k - \mu$. Then,
the optical conductivity can be obtained by calculating the current-current correlation function,
$\Pi(i\omega)$, using the renormalized Green's function obtained from solving Eq.$\!$ (\ref{s1}) 
self-consistently.
The calculated optical conductivity shows a strong reduction of Drude weight 
and a broad MIR absorption, although the MIR feature around 0.06 eV is less pronounced 
and broader compared with experimental observations.

This paper is organized as follows: In the following section, we present the 
Eliashberg-type formalism in the Matsubara frequency
to calculate the renormalized Green's function with the impurity, electron-phonon, and 
Coulomb interactions included self-consistently. 
We then describe the analytic continuation procedure to obtain the renormalization function
$Z(\omega)$ in the real frequency. The optical 
conductivity calculated with the renormalized Green's function is presented in Sec.$\!$ III. 
We will discuss 
how the Drude part and the MIR absorption of $\sigma(\omega)$
are affected
as the impurity scattering rate, the electron-phonon and electron-electron interactions are varied.
These result will then be compare with the experimental observations.
Finally, Sec.$\!$ IV is for the summary and some concluding remarks.

\section{Formalism }
\label{2}
The optical conductivity is calculated from the current-current correlation function, $\Pi(\omega)$, as
$\sigma(\omega) = \frac{i}{\omega} \lim_{q\to 0}\Pi(q,\omega)$ \cite{ma}.
We use the approximation where the electron self energy is momentum
independent. In this case, it can be shown that the vertex correction in the current-current correlation
function vanishes for $ q \rightarrow 0$ \cite{kh}. This leads to
\begin{equation}
\Pi(i\omega_m) \;=\; \frac{2e^2}{3m^2V}\sum_{\vec{p}} \vec{p}^2 \frac{1}{\beta}
\sum_{ip_n} G(\vec{p},ip_n + i\omega_m)G(\vec{p}, ip_n),
\label{pi}
\end{equation}
where $ip_n = \pi T(2n+1)$ and $i\omega_m = 2\pi Tm$ are, respectively, fermion and boson Mastubara frequencies,
where $T$ is the temperature, $m$ and $n$ are the integers.
$\beta = 1/k_B T$, and $V$ is the volume.
The evaluation of Eq.$\!$ (\ref{pi}) using Eq.$\!$ (\ref{s2}) produces 
\ba
\Pi(i\omega)=\frac{2\pi e^2n}{m}\frac{1}{\beta}\sum_{ip}
\frac{\theta(ip+i\omega)-\theta(ip)}{(p+\omega)Z(ip+i\omega)-pZ(ip)}
\ea
in the Mastubara frequency. After performing the analytic continuation of 
$i\omega \to \omega+i\delta$, to the real frequency, the optical conductivity is given by 
\ba
\sigma(\omega) &=&
\frac{1}{\omega} \frac{e^2n}{m} \int_{-\infty}^{\infty} 
d\varepsilon [ f_F (\varepsilon) - f_F (\varepsilon + \omega)]
\qquad\quad\qquad\qquad\qquad\qquad\qquad\qquad \nonumber\\
&\times& Re\Big[ i \frac{\theta(\varepsilon\! +\! i\delta) 
- \theta(\varepsilon\! +\! \omega
+\! i\delta)}{\varepsilon Z(\varepsilon\! +\! i \delta) - (\varepsilon\! +\! \omega)
Z(\varepsilon\! +\! \omega\! +\! i \delta)} 
\Big. \nonumber \\
&& \Big. -i \frac{\theta(\varepsilon\! -\! i \delta)
- \theta(\varepsilon\! +\! \omega\! +\! i \delta)}
{\varepsilon Z(\varepsilon\! -\! i \delta) - (\varepsilon\! +\! \omega)
Z(\varepsilon\! +\! \omega\! +\! i \delta)}\Big] ,
\label{rsigma}
\ea
where $f_F (\varepsilon)=1/(1+e^{\beta\varepsilon})$ is the Fermi distribution function, 
and $\theta(\omega+i\delta) = \tan^{-1}
\big[\frac{i\varepsilon_F}{\omega Z(\omega +i\delta)}\big]$.
The finite conduction bandwidth $B$ with a constant DOS is 
explicitly considered through the factor of $\theta$, which is $\pi/2$
for the usual case of infinite bandwidth metal.
In order to calculate the optical conductivity from Eq.$\!$ (\ref{rsigma}) 
we need $Z(\omega)$ which defines single-particle 
interacting Green's function $G(\omega)$. This can be obtained by solving Eq.$\!$ (\ref{s1})
self-consistently. 
The electron self-energy is obtained by calculating the exchange diagram of 
the renormalized electron Green's function and the effective 
electron-phonon and Coulomb interactions with the vertex correction included 
via the method of Nambu.
The Coulomb interaction, modelled in terms of the onsite Hubbard repulsion for simplicity, 
is included on an equal footing with the 
electron-phonon interaction.
The impurity effects are included with the $t$-matrix approximation.
The Eliashberg-type equation can be written in the Mastubara frequency as
\begin{eqnarray} 
Z_n p_n  &=& p_n + \frac{1}{\beta} \sum_m 
[ \lambda_{ph}(n-m) -\lambda_{ch} (n-m)  \nonumber \\
&+& \lambda_{sp} (n-m) ] 2\theta_m
\Gamma +\frac{1}{\pi \tau} \theta_n ,
\label{eli}
\end{eqnarray}
where 
$\theta_n = \tan^{-1}(\frac{B}{2p_nZ_n})$, and 
$\lambda_{ph}(n-m) = \int_0^{\infty}d\Omega \frac{\alpha^2 F(\Omega)
2\Omega}{[\Omega^2 + (p_n-p_m)^2]}$ is the 
electron-phonon interaction kernel. $\lambda_{ch}(n-m)$ and $\lambda_{sp}(n-m)$ are,
respectively, the interactions in the charge and spin channels due to the 
Hubbard repulsion. They are determined self-consistently as 
\ba
\lambda_{ch}(k) &=& UN_{F} \left\{ \frac{1}{2} - \chi_n + \chi_n^{2}\ln[1+1/\chi_n] \right\},
\label{lch}
\\ 
\lambda_{sp}(k) &=& UN_{F} \left\{ \frac{1}{2} + \chi_n + \chi_n^{2}\ln[1-1/\chi_n] \right\},
\ea
where $\chi_n(k)$ is the dimensionless susceptibility given 
by 
\ba
\chi_n(k) = \frac{N_F U}{\varepsilon_F}\frac{1}{\beta}\sum_{l}\theta_l \theta_{l+k}.
\ea
The $\Gamma$ on the right hand side of Eq.$\!$ (\ref{eli}) represents the vertex correction 
satisfying the Ward-identity\cite{en}.
When we neglect the vertex correction, $\Gamma = 1$. If we assume a weak 
frequency dependence of $\Gamma$,
the vertex function $\Gamma$ reduces to $Z(ip_m)$.
In this work, we treat the vertex correction exactly, and $\Gamma$ 
is given by 
\ba
\Gamma = \Big[\frac{ip_nZ(ip_n) - ip_mZ(ip_m)}{ip_n-ip_m}\Big].
\ea
Solving Eq.$\!$ (\ref{eli}) self-consistently yields $Z(i\omega)$ 
in the Mastubara frequency.
In order to calculate $\sigma(\omega)$, analytic continuation of $i\omega \to \omega + i\delta$
should be performed 
to get $Z(\omega)$ in real frequency.
The numerically exact analytic continuation of standard Eliashberg equation is usually performed 
by the iterative method developed by Marsiglio, Shossmann, and Carbotte (MSC)
using a mixed-representation \cite{mar}.
But when we include the vertex function exactly, the MSC method can not be applied 
because it needs a specific form of equation.
Here, in order to consider vertex correction exactly, we do the alanytic continuation 
by employing the iterative method extended by Takada \cite{ta}.
In this case, Eq.$\!$ (\ref{eli}) is transformed to a mixed representation as fallow:
\ba
Z(\omega) &=&
\tilde{Z} (\omega)
+ \int_{0}^{\infty}d\Omega P(\Omega)
\bigg{\{}[n_B(\Omega) + n_F(\omega + \Omega)]
\nonumber\\
&\times& G(\omega + \Omega)
\Big[\frac{(\omega \!+\! \Omega)Z(\omega \!+\! \Omega)
\!-\!\omega Z(\omega)}{\Omega}\Big]
\nonumber \\
&+& [n_B(\Omega) \!+\! n_F(\Omega \!-\! \omega)] 
G(\omega \!-\! \Omega)
\nonumber\\
&\times& \Big[\frac{(\omega \!-\! \Omega)Z(\omega \!-\! \Omega)
\!-\!\omega Z(\omega)}{-\Omega}\Big]\!\bigg{\}},
\label{elir}
\ea
where 
\ba
\tilde{Z} (\omega)
&=& 1 +\frac{1}{\omega\beta}\sum_{m}\int_0^{\infty}\!\!d\Omega P(\Omega)
\Big(\frac{1}{ip_m\! -\! \omega\! -\! \Omega} - \frac{1}{ip_m\! -\! \omega\! +\! \Omega}\Big)
\nonumber\\
&\times& G(ip_m) \Big[\frac{ip_m Z(ip_m)\! -\! \omega Z(\omega)}{ip_m - \omega}\Big]
+ \frac{i}{\pi \tau}\frac{\theta(\omega)}{\omega},
\nonumber\\
P(\Omega) &=& -\frac{1}{\pi}Im\Lambda(\Omega)\,
\nonumber \\
\Lambda(\Omega) &=& \lambda_{ch}(\Omega)
- \lambda_{ph}(\Omega) - \lambda_{sp}(\Omega)
\nonumber\\
G(ip_m) &=& 2\theta(ip_m), ~
G^{R}(\omega) = 2i\theta(\omega).
\ea
$\tilde{Z} (\omega)$ of Eq.$\!$  (\ref{elir}) represents the renormalization function 
obtained by substituting 
$i\omega$ to $\omega + i\delta$
$before$ the frequency summation.
The second term is the correction to $\tilde{Z} (\omega)$ 
to yield the correct retarded renormalization function $Z(\omega)$
one would have obtained if the analytic continuation were performed $after$
the frequency summation.
Putting the solutions of Eq.$\!$ (\ref{eli}), $Z(i\omega)$, into the $\tilde{Z} (i\omega)$
Eq.$\!$ (\ref{elir}) yields a self-consistent Eliashberg-type equation in the real frequency.
Then, $Z(\omega)$ can be obtained by computing iteratively Eq.$\!$ (\ref{elir}).

In order to model fullerene superconductors, three truncated-Lorentization functions 
were used to represent $\alpha^2 F(\Omega)$ as follow \cite{pro1,pro2}:
\ba
\alpha^2 F(\Omega) &=& \sum_{\nu = 1}^{3} \alpha_{\nu}^{2}F_{\nu}(\Omega),
\nonumber\\
F_{\nu}(\Omega) &=& \left\{ \begin{array}{l}
\frac{1}{R}
\Big[\frac{1}{(\Omega - \omega_{\nu})^2 + \Gamma^2}
- \frac{1}{\Gamma_c^{2} + \Gamma^{2}}\Big], ~
for |\Omega - \omega_{\nu}|\leq \Gamma_c,\\
0,~otherwise,
\end{array} \right.  
\label{alpha}
\ea
where $F_{\nu}(\Omega)$ is the truncated Lorentizian centered at $\omega_{\nu}$ 
with the width of $\Gamma = \omega_{\nu}/5$, 
$\Gamma_c$ is the cutoff frequency of $\Gamma_c = 3\Gamma$,
and $R$ is normalization constant such that $\int_0^{\infty}d\Omega F_{\nu}(\Omega) =1$.
Various theoretical and experimental estimates do not agree well each other 
in terms of distribution of coupling strength $\alpha_{\nu}^2$ among different modes.
These estimates show, however, that the phonon frequency derived from intramolecular 
$A_g$ and $H_g$ modes are distributed over $0.03-0.2$ eV with the total $\lambda$
in the range of $0.5-1$ eV. In view of this, we represent the phonon modes with 
three groups centered around $\omega_{\nu} =0.04,\;0.09,\;0.19$ eV, 
and $2N_F\alpha_{\nu}^2/\omega_{\nu} = 0.3\lambda_s,\;0.2\lambda_s,\;0.5\lambda_s$, 
respectively, for $\nu = 1, 2, 3$. Note that 
$\sum_{\nu = 1}^32N_F \alpha_{\nu}^2/\omega_{\nu}=\lambda_s$. The $\lambda_s$ sets 
the strength of $\alpha^2 F(\Omega)$ and $N_F\alpha^2 F(\Omega)/\lambda_s$ 
is independent of $\lambda_s$. For infinite bandwidth superconductors,
$\lambda$ is equal to $\lambda_s$ in the limit $\Gamma \to 0$. 
For a finite bandwidth system, however, $\lambda$ is reduced from $\lambda_s$
because the available states to and from which quasiparticles can be sccattered 
are restricted as the bandwidth is reduced.

\section{Results}
The self-consistent equation of Eq.$\!$ (\ref{s1}) is solved numerically as described in the 
previous section to obtain $Z(\omega)$. Then, the optical conductivity is calculated 
from the Eq.$\!$ (\ref{rsigma}).
Fig.$\!$ 1 shows the optical conductivity $\sigma(\omega)$ as $\lambda$ is varied 
when the Coulomb interaction $U$ is set to 0 for a reference. 
Here, the Fermi energy $\varepsilon_F$, temperature $T$ and 
impurity scattering rate $1/\tau$ are set to 
$0.25,\; 0.001,\; 0.01$ eV, respectively. This result shows quite a similar behavior 
to van den Brink $et ~al.$ calculation. As $\lambda$ is increased, the width of Drude peak 
becomes narrower and it's weight is transferred to a mid-infrared spectrum. 
However, the reduction of Drude weight is less than the factor of ($1+\lambda$), 
because of the finite bandwidth.
The inset shows a MIR absorption spectra obtained by extracting Drude part from the total 
optical conductivity. In determining the Drude weight, fitting procedure was carefully employed
and confirmed by examining zero frequency extrapolation in the Mastubara frequency 
which is proposed by Scalapino $et\;al$ \cite{sc2}.
The three Lorentizian peaks of $\alpha^2F(\omega)$ in the electron-phonon 
paring kernel are attributted to the development of these MIR peaks.
But, the MIR peaks are broadened and move to slightly high frequencies.
Fig.$\!$ 2 shows the MIR absorption due to the Coulomb interaction. 
The MIR part is also extracted by fitting as shown in the inset. 
In order to focus on how $U$ affects the total optical conductivity, $\lambda$ is set to 0. 
$\varepsilon_F$, $T$ and $1/\tau$ are same as in Fig.$\!$ 1.
The Coulomb interaction induces the strong $\omega$ dependence of 
renormalization function $Z(\omega)$, 
and the low frequency strong $\omega$ dependence of $Z(\omega)$
distorts the Drude part of optical conductivity and induces the MIR absorption 
in the fairly low frequency region.
As the impurity effect is enhanced, the MIR absorption due to Coulomb interactions 
tends to shift to higher frequency and finally
merge together with the MIR peaks developed by electron-phonon interaction,
as shown in Fig.$\!$ 3 for $UN_F = 0.3$ and $\lambda = 0.7$. Note that 
the position of this merged MIR peak in Fig.$\!$ 3 
is around and above $0.06$ eV which is experimentally observed value of DeGiorgi.

Fig.$\!$ 4 is $\sigma(\omega)$ of doped fullerenes with 
$T = 0.005$ eV, $\varepsilon_F = 0.25$ eV, $1/\tau = 0.1$ eV,
$UN_F = 0.3$, and $\lambda = 0.7$, 
which is to be compared with the experimental observations.
The Drude weight
is reduced to 0.467 of the total intra-band optical weight. 
The reduction factor of the Drude weight by electron phonon interaction is 
$1+\lambda$, and the finite bandwidth futher restricts the reduction factor.   
It therefore seems unlike that the Drude weight less than about 0.6 of
the total intra-band spectral weight 
can be explained without the Coulomb interactions, 
when we take $\lambda \approx 0.7 - 0.8$.  
The Coulomb interaction suppresses the Drude part substantially by inducing 
$\omega$ dependence of the renormalization function $Z(\omega)$ 
in the low frequency region.
We think that the large reduction of Drude weight like the DeGiorgi experiment 
is a result of the strong Coulomb interaction between conduction electrons
in addtion to the electron-phonon interaction.
However, our results are still not sufficient to explain experimentally founded results:
(a) The Drude weight is about 0.46 of the total intra-band optical weight
with a resonable set of parameter values while DeGiorgi found $0.1-0.2$.
(b) The MIR absorption is very broad which begins around 0.02 eV, 
has a peak around 0.07 eV and extends well over the Fermi energy.

\section{Summary and Conclusion}
\label{4}
In this paper, we tried to give an explanation for the unusual behavior of 
optical conductivity in the normal state $\rm{A_3 C_{60}}$.
It is generally accepted that the fullerene superconductor could be 
characterized by the phonon-mediated $s$-wave superconductor
\cite{degiorgi,gu2}.
However, a few experiments like optical conductivity still remain not understood by 
the electron-phonon scattering together with the disorder effects.
Our motivation lies in that the fullerene superconductors have such a 
narrow bandwidth that the phonon frequency, the Coulomb interaction, and 
the Fermi energy are all comparable, $\omega_{ph} \sim V \sim \varepsilon_F$.
In order to consider properly the Coulomb interaction and 
the narrow bandwidth of fullerene superconductors,
the self-consistent Eliashberg-type coupled equations are solved to obtain 
the renormalized Green's function.
The theory includes the frequency dependent screened Coulomb interaction
together with the electron-phonon interaction and includes the vertex correction via 
Nambu's method. In order to treat the vertex function exactly, analytic continuation 
is performed via the iterative method of mixed repersentation which is developed by Takada.
Once we get renormalization function $Z(\omega)$ in real frequency, we can calculate 
optical conductivity in normal states.
As we expected, the electron-phonon interaction is not suffficient to resolve 
the substantial reduction of Drude weight and pronounced MIR peak.
The strong Coulomb interaction induces $\omega$ dependence in renormalization 
function $Z(\omega)$. As a result, the Drude form in optical spectra is 
distorted accompanying the reduction of Drude weight.
When the impurity effect is enhanced, the MIR absopption induced by strong Coulomb interaction
merge together with the MIR peaks due to electron-phonon scattering showing large reduction 
of Drude weight and MIR peak around $0.06$ eV. 
Although it is not sufficent to explain experimentally founded 
results, our result is close to DeGiorgi's experiment.
We improve Gunnarsonn's calculation by considering the electron-electron interaction and 
finite bandwidth effects explicitly.
In conclusion, the unusual behavior of optical conductivity of the normal state $\rm{A_3C_{60}}$
reveals the fact that both the Coulomb interaction and electron-phonon interaction are important
in examining dynamical properties of fullerene superconductors.


\vspace{3cm}

Figure Captions

\vspace{0.5cm}
Figure 1.
The optical conductivity as a function of 
$\omega$ for various electron-phonon coupling 
constants $\lambda$ when $T=0.001$ eV, $\varepsilon_F=0.25$ eV and $1/\tau=0.01$ eV. 
$U$ is set to 0 for a referrence. As $\lambda$ is increased, 
the width and the weight of the Drude peak
are reduced. The inset shows a decomposition of the total conductivity 
into the Drude and MIR parts for $\lambda = 0.7$.

\vspace{0.5cm}
Figure 2.
The MIR spectra induced by the Coulomb interaction when 
$T=0.001$ eV, $\varepsilon_F=0.25$ eV, $1/\tau=0.01$ eV and $\lambda=0$.
The inset shows a decomposition of the total conductivity, as in Fig.$\!$ 1, 
into the Drude and MIR parts for $UN_F = 0.5$.

\vspace{0.5cm}
Figure 3.
The MIR spectra as the impurity scattering rates $1/\tau$ are varied
when $T=0.005$ eV, $\varepsilon_F=0.25$ eV,
$\lambda=0.7$ and $UN_F=0.3$.
When $1/\tau=0.01$ eV the lower peak is mainly from the Coulomb interaction
while the other peaks are from the electron-phonon interaction.
As $1/\tau$ is increased, these peaks are merged altogether and finally evolve into 
a single broad peak around $0.06 \sim 0.1$ eV.

\vspace{0.5cm}
Figure 4.
The total optical conductivity with it's Drude and MIR parts 
for $T=0.005$ eV, $\varepsilon_F=0.25$ eV, $1/\tau=0.1$ eV, $\lambda=0.7$ and $UN_F=0.3$.
The Drude part in the low frequency region is substantially suppressed 
due to the strong Coulomb interaction.
Consequently, the missing spectral weight is transfered to the broad MIR peak, 
which peaks around 0.07 eV and extends well into the higher energy region.
The ratio of the MIR spectral weight to the total intra-band spectral weight
is 0.533.

\vspace{-0cm}
\begin{figure}
\center
\epsfig{figure=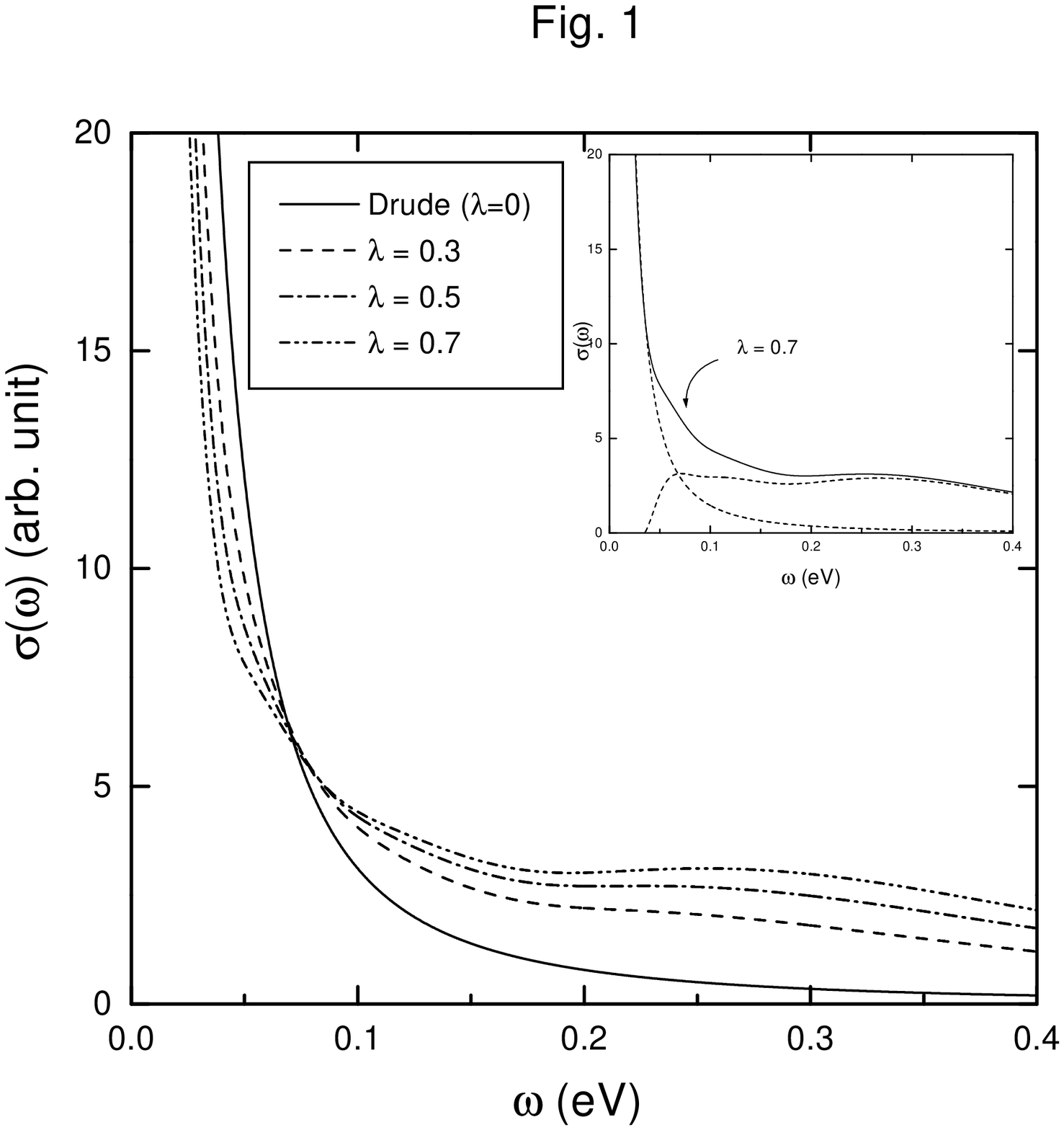,width=0.9\linewidth}
\caption{}
\end{figure}

\newpage
\vspace{-30cm}
\begin{figure}
\center
\epsfig{figure=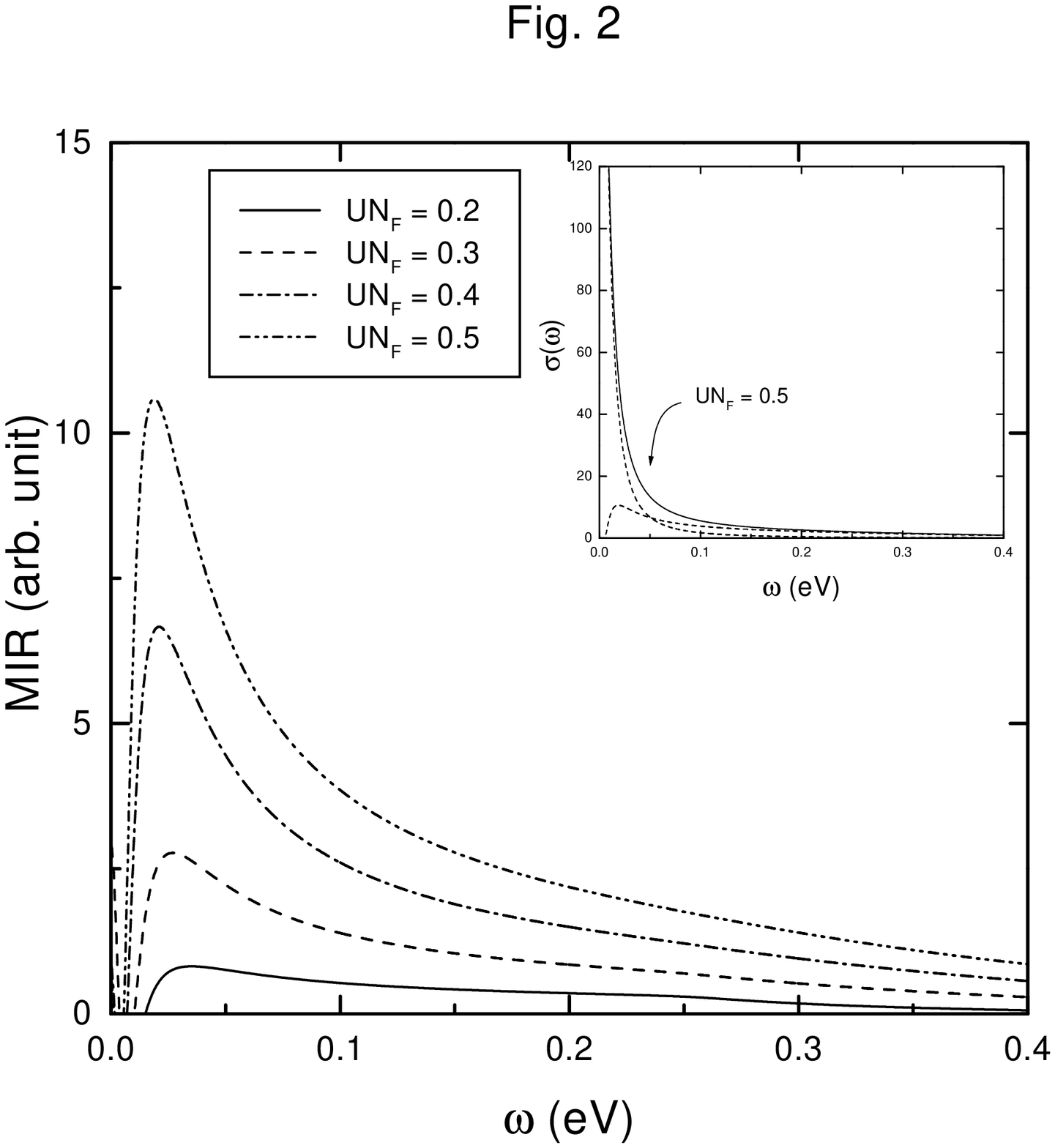,width=0.9\linewidth}
\caption{}
\end{figure}

\vspace{-2cm}
\begin{figure}
\center
\epsfig{figure=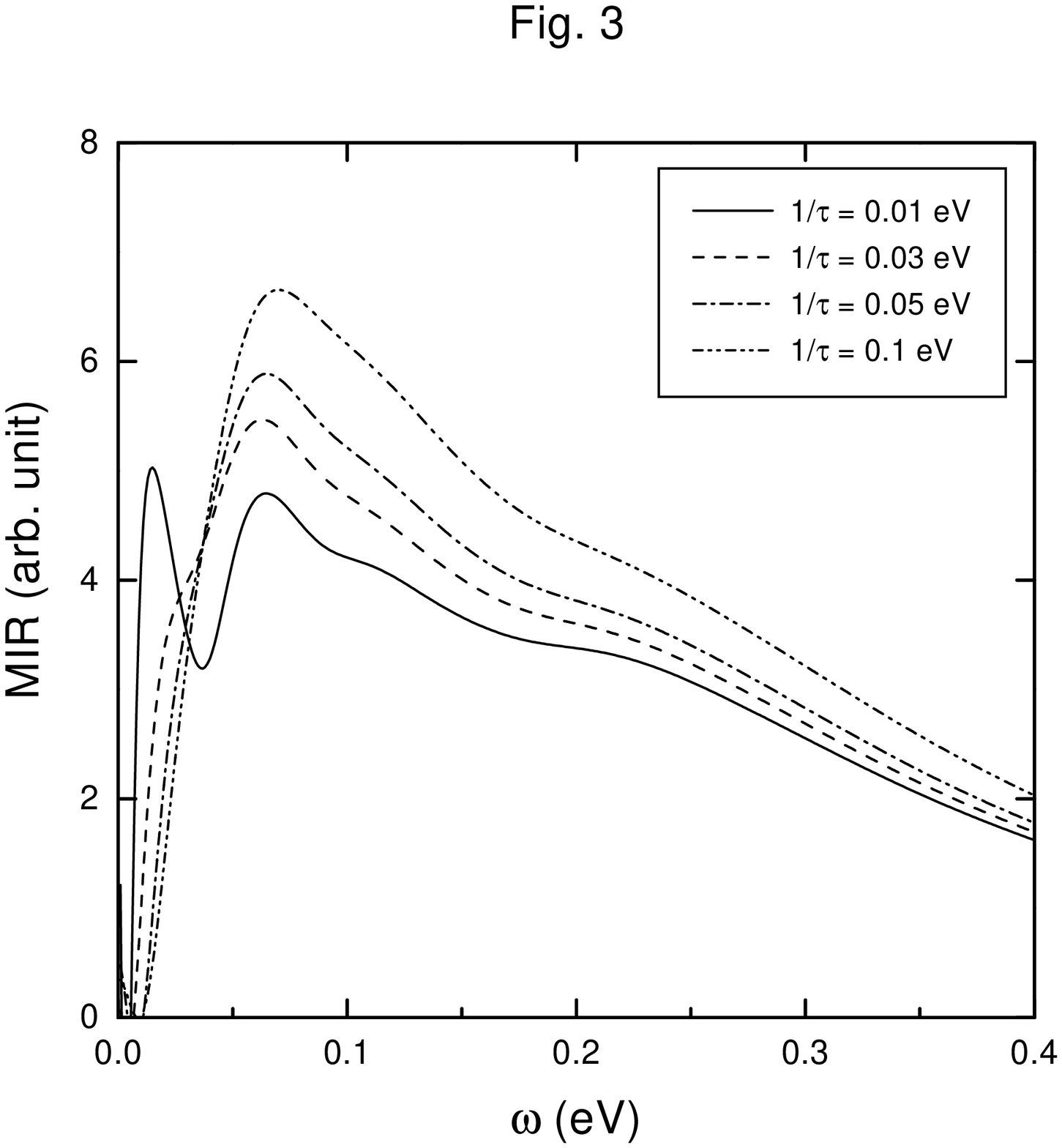,width=0.9\linewidth}
\caption{}
\end{figure}

\vspace{-2cm}
\begin{figure}
\center
\epsfig{figure=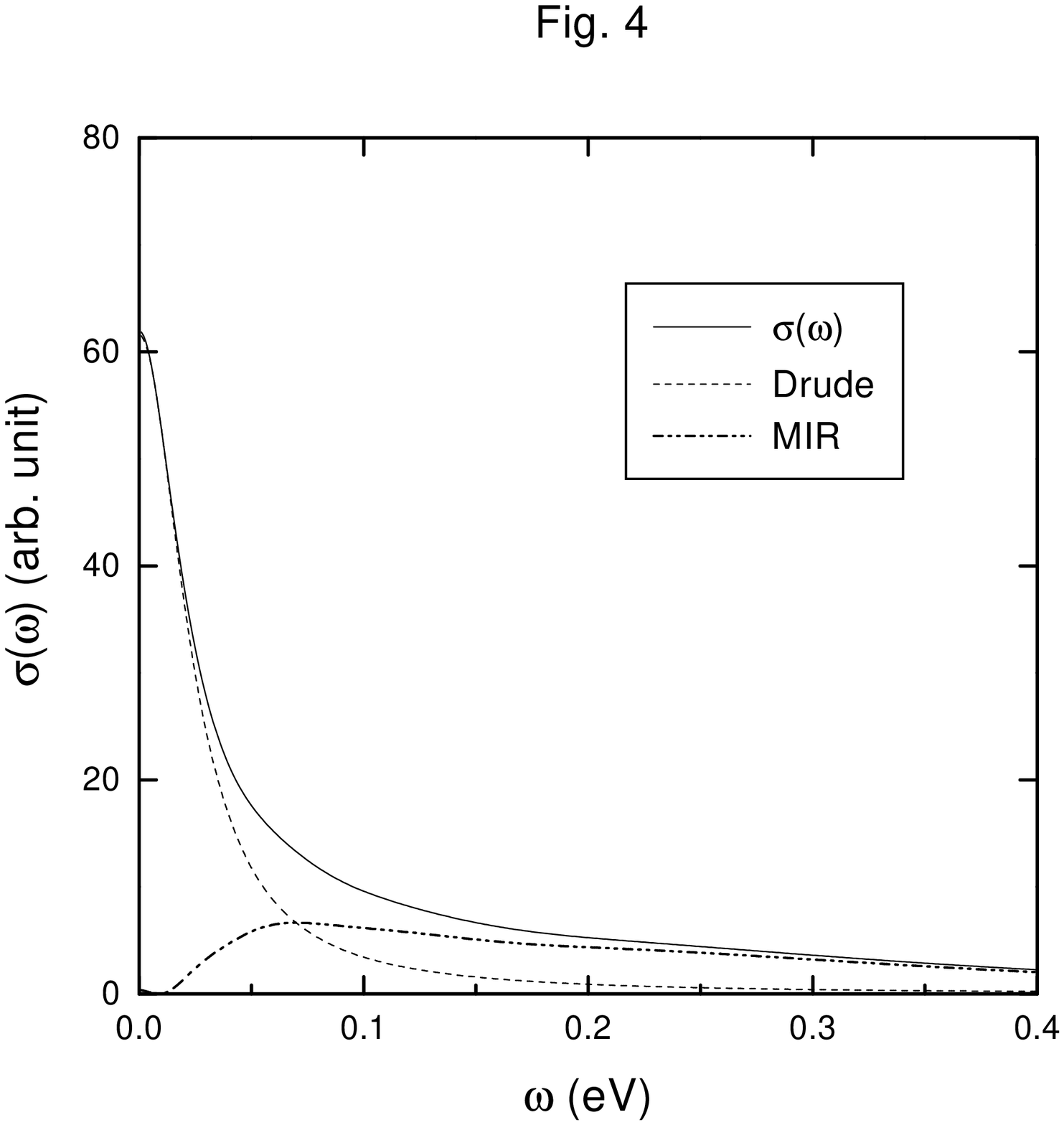,width=0.9\linewidth}
\caption{}
\end{figure}

\end{multicols}{2}

\end{document}